\documentclass[twocolumn,
nofootinbib,
 amsmath,amssymb,
 aps,
 prd,
 floatfix,
]{revtex4}
\usepackage{graphicx}
\usepackage{dcolumn}
\usepackage{bm}
\usepackage{xcolor}
\usepackage[center]{caption}
\usepackage[normalem]{ulem}

\usepackage{hyperref}

\newcommand{\vEDM}{\vec E_\mathrm{DM}}

\newcommand{\Tobs}{T_\mathrm{obs}}
\newcommand{\RIN}{\mathrm{RIN}}
\begin{document}

\title{Search for vector dark matter in microwave cavities with Rydberg atoms}

\author{Jordan Gu\'e,\footnote{jordan.gue@obspm.fr} Aur\'elien Hees, J\'er\^ome Lodewyck, Rodolphe Le Targat, Peter Wolf}
\affiliation{%
 SYRTE, Observatoire de Paris, Universit\'e PSL, CNRS, Sorbonne Universit\'e, LNE, 61 avenue de l’Observatoire 75014 Paris, France\\
}%
\date{\today}

\begin{abstract}
We propose a novel experiment to search for dark matter, based on the application of an electric field inside a microwave cavity and electrometry using Rydberg atoms. We show that this kind of experiment could be extremely useful for detecting specific dark matter candidates, namely massive vector fields coupled to the photon field, more commonly known as dark photons. Such a massive vector field is a good candidate for dark matter. 
Using realistic experimental parameters we show that such an experiment could improve the current constraint on the coupling constant of the dark photons to Standard Model photons in the 1 $\mu$eV to 10~$\mu$eV mass range, with the possibility of tuning the maximum sensitivity via the cavity size. The main limiting factors on the sensitivity of the experiment are the amplitude stability of the applied field and the measurement uncertainty of the electric field by the atoms.
\end{abstract}

\maketitle


\section{\label{sec:Intro}Introduction}

While required to explain several astrophysical and cosmological observations, the microscopic nature of dark matter (DM) is still to this day one of the biggest mysteries in physics \cite{bertone:2018aa}. Among many other classes of DM, ultra-light dark matter (ULDM) models have recently gained a lot of attention in the scientific community, due to the absence of signals from historically dominant models, such as weakly interacting massive particles (WIMPs). These models are characterized by particles with low mass, from $10^{-22}$ eV to $0.1$ eV, meaning that detection from recoiling in particle accelerators is difficult or impossible. Thus, other kind of experiments have to be considered for their detection.

One particularly well motivated model of ULDM is called the dark photon (DP), a spin-1 field of mass $m$ which appears in many beyond the Standard Model theories.  Cosmologically speaking, the DP field is frozen after inflation and starts to oscillate at its Compton frequency when $m c^2/\hbar \gg H$. It can be shown that these oscillations scale as $a^{-3}(t)$,  at late cosmological times, behaving as cold dark matter (CDM). The production of CDM in this case is done through non-thermal processes, such as the so-called misalignment mechanism. 

This additional $U(1)$ field is also well motivated on the particle physics side, as the DP model is a particular case of the so-called $U$ boson, which makes the minimal gauged Standard Model extension, through its coupling with the B-L current of the Standard Model \cite{U-boson}.

For these reasons, DP is a very good light dark matter candidate, with many new experiments being developed to hunt for it. In the following, we will be interested in the coupling of the DP field with the electromagnetic (EM) field with strength $\chi$. This coupling induces an electric field filling all space, with amplitude directly proportional to $\chi$.
In this paper, we propose a new kind of experiment aiming at detecting DP, through this additional electric field. The experiment is based on the use of a microwave cavity in which, an applied external electric field acts as an amplifier for the weak DP induced electric field, with the strength of amplification directly proportional to the square root of the injected power. If these two different electric fields have close frequencies the total field power will oscillate at their frequency difference $\Delta\omega$ and become detectable by atoms located at the center of the cavity through the Stark effect.
We show that using this setup, one could scan new regions of the m-$\chi$ parameter space, in particular in the 1~$\mu$eV to~10~$\mu$eV mass range, with the possibility of tuning the maximum sensitivity regions via the cavity size and design. 
We use some simplifying assumptions, in particular for the model of the cavity. The main aim of our paper is to obtain a rough estimate of the experimental sensitivity using realistic parameters and taking into account the main expected noise sources and systematic effects. Our results and described methods will be useful for the design of a real-life experiment, in which case more careful modeling should be carried out. Nonetheless, we do not expect that to change the orders of magnitude of the sensitivity presented here.


\section{Field equations}
The Lagrangian describing the interaction between the EM field $A^\mu$ and a DP field $\phi^\mu$ of mass $m$ is given by\footnote{In this manuscript, we use the metric signature $(-,+,+,+)$ and the unit of the DP field $\phi^\mu$ is Vs/m, as the usual EM vector potential $A^\mu$.} \cite{Holdom86}
\begin{align}\label{lagrangian}
  \mathcal L =& - \frac{1}{4\mu_0}F^{\mu\nu}F_{\mu\nu}+j^\mu A_\mu  - \frac{1}{4\mu_0}\phi^{\mu\nu}\phi_{\mu\nu} \\ \nonumber
  &\qquad - \frac{m^2c^2}{2\mu_0\hbar^2}\phi^\mu\phi_\mu -\frac{\chi}{2\mu_0}F_{\mu\nu}\phi^{\mu\nu}\, ,
\end{align}
where $F_{\mu\nu}=\partial_\mu A_\nu - \partial_\nu A_\mu$ is the usual electromagnetic field strength tensor, $\phi_{\mu\nu}=\partial_\mu \phi_\nu - \partial_\nu \phi_\mu$ is the DP field strength tensor, $\chi$ is the dimensionless kinetic mixing coupling parameter which characterizes the coupling between the DP and the EM field, and $j^\mu$ is the usual electromagnetic 4-current. A change in the EM 4-potential 
\begin{equation}\label{eq:barA}
	\bar A^\mu=A^\mu+\chi \phi^\mu
\end{equation}  allows one to redefine the fields in terms of mass eigenstates called massless and massive photons (at the price of a non-standard interaction between the EM field and standard matter), see the discussion in \cite{nelson:2011tv}. Using this change of variable, Eq.~\eqref{lagrangian} becomes, at first order in $\chi$,
\begin{align}
	\mathcal L =& - \frac{1}{4\mu_0}\bar F^{\mu\nu} \bar  F_{\mu\nu}+j^\mu \left(\bar A_\mu -\chi \phi_\mu\right)  - \frac{1}{4\mu_0}\phi^{\mu\nu}\phi_{\mu\nu} \\ 
  &\qquad - \frac{m^2c^2}{2\mu_0\hbar^2}\phi^\mu\phi_\mu \nonumber \, ,
\end{align}
with $ F_{\mu\nu} = \bar F_{\mu\nu}-\chi \phi_{\mu\nu}$.  The field equations read
\begin{subequations}\label{eq:field_eq}
  \begin{align}
    \bar F^{\beta\alpha}_{\, , \alpha} &= \mu_0 j^\beta \, , \\
    \phi^{\beta\alpha}_{\, , \alpha}   &= -\chi \mu_0 j^\beta - \frac{m^2c^2}{\hbar^2} \phi^\beta \, .
  \end{align}
The antisymmetry of both strength tensors leads to the conservation of the electromagnetic 4-current $\partial_\mu j^\mu = 0$,  and the continuity equation for the DP field $\partial_\mu \phi^\mu = 0$. 
\end{subequations}
Using the Lorenz gauge for EM, $\partial^\mu \bar A_\mu = 0$, Eqs.~(\ref{eq:field_eq}) become 
  \begin{subequations}\label{eq:field_mass}
    \begin{align}
      \Box \bar A^\beta &= -\mu_0 j^\beta \, , \\
      \Box \phi^\beta  &= \frac{m^2c^2}{\hbar^2} \phi^\beta + \chi \mu_0 j^\beta \, , \label{eq:k2m}
    \end{align} 
  \end{subequations}    
where $\Box = \eta^{\mu\nu}\partial_\mu \partial_\nu \equiv -\frac{1}{c^2}\partial^2_t + \nabla^2$. These equations
admit two classes of solutions in vacuum: a massless vector field (standard EM) and a massive one,  the latter being characterized by solutions : 
\begin{subequations}
\begin{align}
	\phi^\beta &=Y^\beta e^{i k_\mu x^\mu}\,   \label{eq:osc_DP} , \\
	\bar A^\alpha &= 0 \, ,
\end{align}
\end{subequations} 
where 
\begin{equation}
    k_\mu k^\mu = -\frac{\omega^2}{c^2} + \left|\vec k\right|^2 = -\left(\frac{mc}{\hbar}\right)^2.\label{eq:dispersion}
\end{equation}
As can be noticed from Eq.~(\ref{eq:barA}), this solution induces an ordinary electromagnetic field in a vacuum (see also \cite{Horns})
\begin{equation}\label{eq:A}
	A^\beta = -\chi Y^\beta e^{i k_\mu x^\mu}\, . 
\end{equation}
The stress-energy tensor from the DP field can be derived from Eq.~(\ref{lagrangian}) and reads
\begin{align}
	  T_{\mu\nu} &= \frac{1}{\mu_0}\phi_{\mu\alpha}\phi_\nu^{\ \alpha}-\frac{1}{4\mu_0}\eta_{\mu\nu}\phi^{\alpha\beta}\phi_{\alpha\beta} \\
	  &- \frac{m^2c^2}{\mu_0\hbar^2}\left(\frac{1}{2}\phi^\alpha \phi_\alpha \eta_{\mu\nu}-\phi_\mu \phi_\nu\right)\, .\nonumber
\end{align}
For the plane-wave solution derived in Eq.~(\ref{eq:osc_DP}), the time-averaged value of the stress-energy tensor is given by   
\begin{align}
  \langle T_{\mu\nu} \rangle& = \frac{1}{2\mu_0}k_\mu k_\nu Y_\alpha Y^\alpha \, ,
\end{align}
which leads to an estimate of the energy density 
\begin{equation}\label{energy_density}
  \rho = \langle T_{00}\rangle= \frac{\omega^2 |\vec Y|^2}{2\mu_0c^2}\, ,
\end{equation} 
for $\vec k=0$. Similarly, the averaged pressure vanishes. This shows that the DP can indeed be interpreted as cold and pressureless DM and, under this assumption, the amplitude of oscillation of this vector field, i.e. $\vec Y$, depends directly on the local DM energy density $\rho_{\mathrm{DM}}$. Its experimental value has been considered for a while to be $\rho_{\mathrm{DM}} =$ 0.4  GeV/cm$^3$ \cite{McMillan}, but has recently been readjusted such that $\rho_{\mathrm{DM}} \in \left[0.3 ; 0.55\right]$ GeV/cm$^3$ \cite{GaiaSausage}.

In conclusion, in vacuum, a DP will behave as a plane wave whose dispersion relation is given by Eq.~(\ref{eq:dispersion}). On average, this DP will behave as cold DM and its local energy density is directly related to the amplitude of the oscillation and to its mass. Due to the coupling between the DP and the EM field (coupling characterized by the mixing parameter $\chi$), the DP field will induce a small electromagnetic field, whose strength is proportional to $\chi$ and to the DP field amplitude, see Eq.~(\ref{eq:A}). In particular, if one considers that $\vec k=0$, (which is justified to leading order in $v_\mathrm{DM}\sim 10^{-3}c$, where $v_\mathrm{DM}$ is the typical DM galactic velocity, see Appendix~\ref{app:static_DP}), then the induced electromagnetic field consists mostly in an oscillating electric field the magnetic field component is suppressed by a factor $v_\mathrm{DM}/c$.) of the form \cite{Horns}
\begin{subequations}\label{eqs:EDM}
\begin{equation}
	 E^j_\mathrm{DM} = -\frac{\partial  A^j}{\partial t} = -i \chi \omega  Y^j e^{-i \omega t}\, .
\end{equation}
The amplitude of this oscillating electric field is therefore directly related to the local DM density through
\begin{equation}
	\left|\vEDM\right| = \chi c\sqrt{2\mu_0 \rho_{\mathrm{DM}}}\, .
 \label{amp_E_field}
\end{equation}
\end{subequations}
The idea of several experiments searching for DP is to focus this small electromagnetic field in order to enhance it and hopefully make it detectable \cite{SHUKET,Tokyo1,Tokyo2,Tokyo3,Tokyo4, FUNK}.

\section{\label{sec:Modeling_exp}Theoretical modeling of the experiment}

As mentioned in the previous section, an oscillating DP coupled to the standard electromagnetic field will induce a small oscillating electric field in a vacuum. One very peculiar feature of this electric field is that it does not propagate, i.e. its wave vector vanishes $\vec k=0$ (to first order in $v_\mathrm{DM}/c$). This is due to the fact that this electric field is induced by a massive vector field and therefore its dispersion relation is given by Eq.~(\ref{eq:dispersion}). In this section, we will show how an electromagnetic cavity can be used to search for the electric field induced by a DP. In addition, we will show how to use atoms as a tool to detect this electric field through the Stark effect, i.e. the displacement of the energy levels of an atom under a perturbation by a static electric field (or by an electric field whose frequency is much lower than the transition frequency of the atom). There are mainly two reasons to consider a cavity as an experiment to search for DP.

First of all, as for other DP experiments using resonators, the mirrors of the cavity will enhance the electric field induced by the DP. Indeed, the electric field parallel to the surface of a perfect conductor has to vanish. Therefore, because of the presence of the oscillating DP-induced electric field, the mirror will generate a standard electromagnetic field that will propagate perpendicularly to the mirror and whose amplitude is such that it will cancel the DP-induced field parallel to the surface. Physically, the DP-induced electric field will induce an oscillation of the electrons within the mirror which will create a standard electromagnetic field. Since a cavity consists in two mirrors, this boundary conditions can, under some conditions, produce resonances that will significantly enhance the small DP-induced electric field. 

The second reason  to consider a cavity is related to the use of atoms to measure the electric field inside the cavity through the Stark effect, which is sensitive to the square of the electric field. If one applies a standard electromagnetic wave inside the cavity (whose electric field will be denoted by $\vec E_a$) the DP contribution to the square of the electric field inside the cavity is $\sim \vec E_\mathrm{DM}\cdot \vec E_a$ (to first order in $\chi$), which can also be enhanced by a resonant $\vec E_a$. In addition to enhancing the amplitude of the signal to be measured, applying an external field is also important to produce a signal at low, but non-zero, frequency, where the Stark effect can be realistically measured. More precisely, for a cavity whose length is of the order of a few cm, we will be interested in searching for DP oscillating at a frequency $f=\omega/2\pi$ of the order of a few GHz. 
The difficulty is that such a rapid oscillation of the atomic transition frequency will be very hard to measure as the interrogation cycle of the atoms is much longer. But, if one applies en external field at angular frequency $\omega_a$ which is close to $\omega$, then the cross term between the DP electric field and the applied electric field will have a component oscillating at the low angular frequency $\left|\omega-\omega_a\right|$, which can be measured by the atoms.

\subsection{Expression of the total electric field inside the cavity}\label{sec:model_E}

In this section, we derive the electric field induced by both the DP and the applied field at the center of the cavity. Here, we only summarize the methodology and discuss the main results. The detailed derivation can be found in Appendix~\ref{Fields_expression}. We consider a cavity consisting of two flat mirrors of reflectivity $r$ separated by a distance $L$ \footnote{For simplicity we assume that the transverse size of the mirrors is $\gg L,\lambda$, where $\lambda$ is the wavelength of the fields of interest.}. The reflectivity and the cavity quality factor $Q$ are related through (Q $\gg$ 1)\cite{Suter14}
\begin{equation}
	r \approx 1 - \frac{\pi}{2Q} \, .
\end{equation}
where we considered only low resonance modes, such that $Q \sim \mathcal{F}$, the finesse of the cavity.
The principle of calculation is similar for both the DP and the applied electric fields and can be found in \cite{savallePhD}. The idea is to propagate the electric field an infinite number of round trips inside the cavity and to sum these infinite number of contributions at a given location. To perform cavity round trips, the field is propagated along one direction and when it reaches a mirror, its amplitude is multiplied by $-r$ and its wave vector is flipped. For $r<1$, the infinite sum converges and can be calculated explicitly.

\begin{figure}[hb!]
\centering
\includegraphics[scale=0.35]{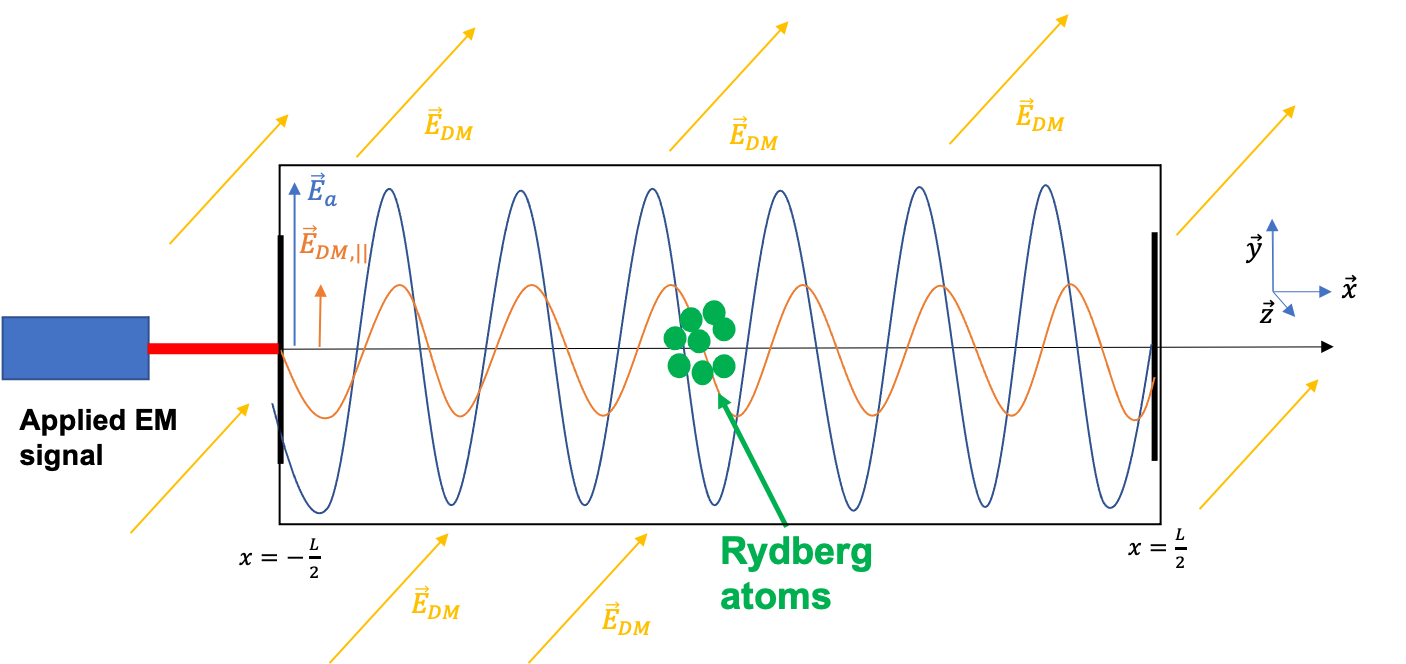}
\caption{An external field (in blue) is applied at the cavity edge. The standing DP electric field (in yellow) generates a propagating electric field inside the cavity (in orange). At the center of the cavity, the transition frequency of Rydberg atoms is impacted by $|E|^2$ through the Stark effect (see text.).}
\label{DP_cavity}
\end{figure}

First, let us apply this procedure to the applied external field. We  assume that the external electric field is applied on the left edge of the cavity, see Fig.~\ref{DP_cavity}. After entering the cavity, the wave will undergo an infinite number of round trips inside the cavity and it will lead to the first contribution to the total field, see Fig. \ref{DP_cavity}.
Ideally, the applied field can be parametrized as $\vec E_{a}=\vec X_a \Re \left[  e^{-i(\omega_at - k_a (x+\frac{L}{2})+\phi})\right]$, with its amplitude $X_a$, angular frequency $\omega_a$ and phase $\phi$.
Applying the mathematical procedure described above (see Appendix~\ref{ap:applied_field}), the expression of the total applied field at the center of the cavity is given by
\begin{align}\label{eq:E_tot_a}
\vec E^\mathrm{tot}_a = \vec A(\omega_a)\cos(\omega_a t +\phi) + \vec B(\omega_a)\sin(\omega_a t + \phi)\, , 
\end{align}
where $\vec A(\omega_a), \vec B(\omega_a)$ are functions (given by Eqs.~(\ref{eq:A_B})) depending on $X_a$, $r$ and $L$, respectively the applied field amplitude, the reflectivity and the length of the cavity.

Let us now focus on the contribution from the DP field. From Eqs.~(\ref{eqs:EDM}), one can, without loss of generality, write the expression of the electric field related to the DP as $\vec E_\mathrm{DM} = \vec X_\mathrm{DM} \Re \left[ e^{-i\omega t}\right]= X_\mathrm{DM} \hat e_\mathrm{DM} \Re \left[ e^{-i\omega t}\right]$ where $\hat e_\mathrm{DM}$ is a unit vector characterizing the polarization of the DP field and $X_\mathrm{DM}=\chi c\sqrt{2\mu_0\rho_\mathrm{DM}}$. Because of this electric field, both mirrors will generate a propagating standard electromagnetic field such that the total component of the electric field parallel of the mirrors' surface vanishes. These two electromagnetic waves will follow an infinite number of round trips inside the cavity, loosing some energy at each reflection with reflection coefficient $r<1$.  The calculation regarding this contribution is detailed in Appendix~\ref{Fields_expression} and the resulting total electric field induced by the DP at the center of the cavity reads
\begin{align}
\vec E^\mathrm{tot}_\mathrm{DM} = \vec C(\omega)\cos(\omega t) + \vec D(\omega)\sin(\omega t)\, ,
\end{align}
where $\vec C(\omega),\vec D(\omega)$ are functions (given by Eqs.~(\ref{eq:C_D})) of $\vec X_\mathrm{DM}$, $\vec X_\mathrm{DM, \parallel}$, $r$ and $L$ with the first two parameters being respectively the full amplitude of the DP field and its component parallel to the cavity plates, both linearly dependent on the coupling $\chi$. 

The calculations presented in this section have been carried out to leading order in $\mathcal O(\chi)$. In particular, at each interaction between the EM waves and the mirrors a small quantity of EM energy will be transformed into DP. The amplitude of such a process is proportional to $\chi$ and therefore neglected as it contributes terms of order $\mathcal O(\chi^2)$. Furthermore the corresponding energy loss is much smaller than the one coming from the finite reflection coefficient $r$. 

\subsection{Modeling of the Stark effect}

As mentioned in the introduction of this section, the main idea of the experiment proposed here is to detect the hypothetical electric field induced by the DP field by using atoms to measure it through the Stark effect. The Stark effect consists in a shift in the energy levels of an atom under the perturbation of a static (or slowly evolving\footnote{As long as the angular frequency of oscillation is much smaller than the atomic transition angular frequency from state k to i, $\Delta \omega << \omega_{ik}$.}) electric field, and is given by \cite{cohen-tannoudji:1986aa}
\begin{equation}\label{stark}
\Delta \nu = -\frac{\Delta \alpha}{2h} \left| \vec E \right|^2  \, ,
\end{equation}
where $h$ is the Planck constant, $\Delta \nu$ is the frequency shift induced by the slowly evolving electric field $\vec E$ and $\Delta \alpha$ is the differential polarizability of the atomic transition considered. 

Taking into account both contributions from the applied electric field and the DP field computed in the previous section, the total electric field power at the center of the cavity is

\begin{widetext}
\begin{align}
|\vec E(\omega,\omega_a)|^2 &=  \left|\vec E^\mathrm{tot}_a +\vec E^\mathrm{tot}_\mathrm{DM}\right|^2 \,\nonumber\\ 
&=\left(\vec A(\omega_a)\cdot \vec C(\omega)+\vec B(\omega_a)\cdot \vec D(\omega)\right)\cos(\Delta\omega t+\phi)+\left(\vec B(\omega_a)\cdot \vec C(\omega)-\vec A(\omega_a)\cdot \vec D(\omega)\right)\sin(\Delta\omega t+\phi) \label{general_E_power}\\
&+ \mathrm{constant \ and  \ fast \ oscillating \ terms} \, , \nonumber
\end{align}
with $\Delta \omega = \omega_a - \omega$. In the following, we will not consider the constant terms. Indeed, in the experimental scheme proposed here, we will be interested in the oscillatory behavior of the atomic frequencies. Moreover, we discarded the fast oscillating terms whose angular frequencies are $2\omega_a$, $2\omega$ or $\omega_a+\omega$, with periods much shorter than the atom interrogation time, such that on average, their impact vanishes.

Inserting the expressions of $A$, $B$, $C$ and $D$ from Eqs.~(\ref{eq:A_B}) and (\ref{eq:C_D}) and using the Eq.~\eqref{amp_E_field}, the amplitude of the oscillation can be simply written in the form (See Appendix \ref{ap:signal_amp})

\begin{equation}
    \label{eq:Etot2}
     \frac{\chi \beta c\sqrt{1-r^2}\sqrt{2\mu_0\rho_\mathrm{DM}} X_a}{\sqrt{1+2r\cos\left(\frac{\omega_a L}{c}\right)+r^2}}\sqrt{1+4\frac{1+(1+r)\cos\left(\frac{\omega L}{2c}\right)}{1+2r\cos\left(\frac{\omega L}{c}\right)+r^2}} \equiv \chi S(\omega,\omega_a;\rho_\mathrm{DM}, X_a;L,r) \, ,
\end{equation}
\end{widetext}
where 
\begin{equation}
    \beta = \hat e_\mathrm{DM} \cdot \frac{\vec X_a}{X_a} \, ,
\end{equation}
is the projection of the polarization of the DP field on the polarization of the injected electric field. If the polarization is fixed and does not change each coherence time, $\beta = \cos\theta$, with $\theta$ the angle between $\vec X_a$ and $\vec X_\mathrm{DM}$. If the DP field is isotropically distributed $ \beta=1/\sqrt{3}$.  To avoid any orthogonality between the two polarizations, i.e $\beta =0$, one must run the experiment for a significant time, i.e at least several days (see discussion in \cite{Caputo})\footnote{Note that the signal Eq.~\eqref{eq:Etot2} is the largest when $X_a$ is maximum at the center of the cavity, which favours TE01 or TE10 modes. }.
\begin{figure*}[ht!]
    \centering
    \captionsetup{justification=centering}
     \includegraphics[width=0.7\linewidth]{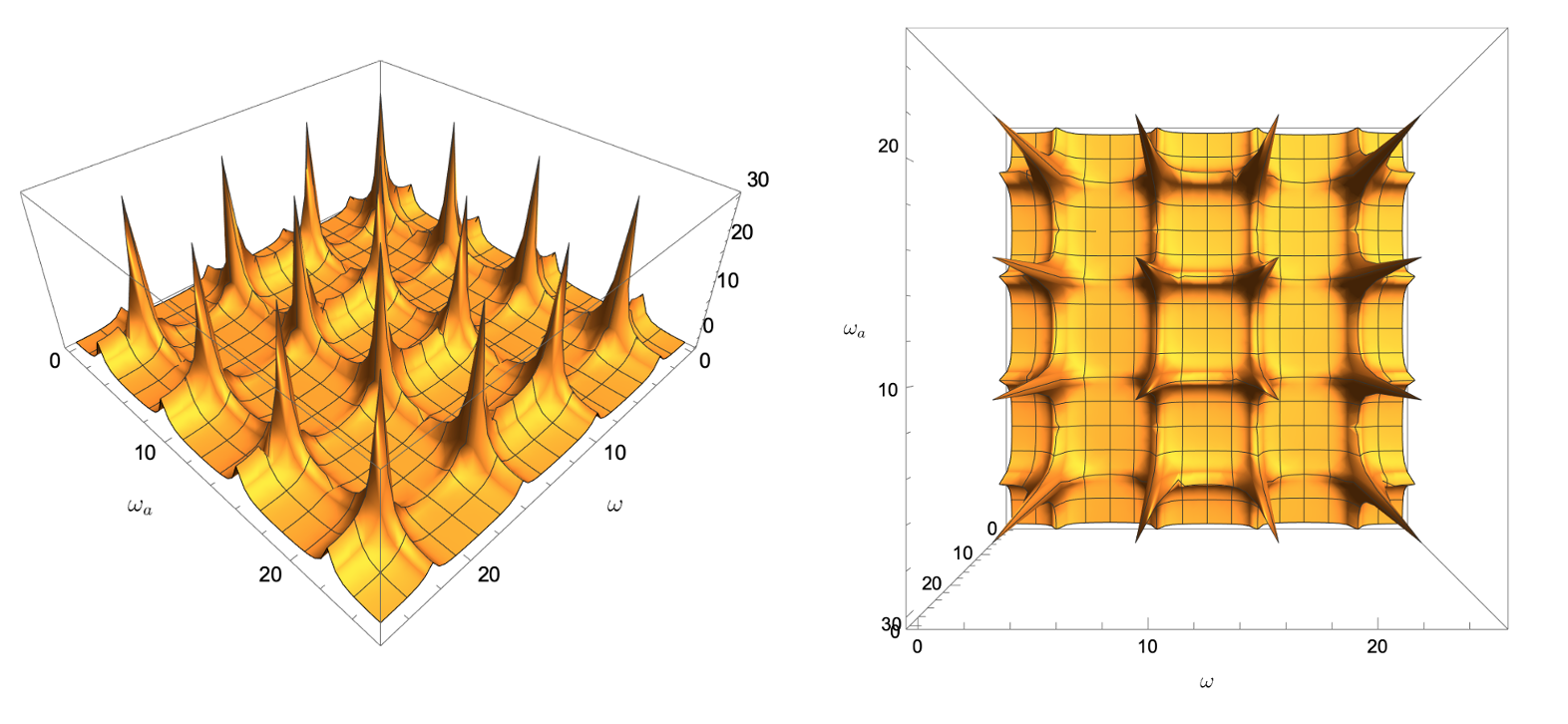}
    \caption{3D plots (left: side view ; right: top view) of the signal contribution Eq.~\eqref{eq:Etot2} (arbitrary units) as function of $\omega, \omega_a$,  in units of $c/L$. Resonance peaks appear clearly when both frequencies correspond to odd modes of the cavity.}
    \label{3D_signal}
\end{figure*}
In Fig.~\ref{3D_signal}, 3D plots of the signal contribution Eq.~\eqref{eq:Etot2} are shown as a function of the angular frequencies of the DM field $\omega$ and the applied field $\omega_a$. One can notice the various resonance peaks when both frequencies correspond to an odd mode of the cavity. In the experiment scheme presented here, we require the two frequencies to be close, hence we are only interested in regions where $\omega \sim \omega_a$. This corresponds to the diagonal of the plot seen from the top. However, in other types of experiment where both frequencies could be separated by some hundreds of MHz or more, other regions of this plot could be studied. 

The amplitude of the signal is linear in $\chi$ and depends non-linearly on the local DM density, on the amplitude and frequency of the applied external field and on the mass ($mc^2=\hbar \omega$) of the DP field. It depends also on some geometric factor of order unity that characterizes the DP field polarization.

For DM and applied frequencies close to odd resonances of the cavity, i.e $\omega L/c \approx \omega_a L/c \approx \pi + 2n\pi$, with $n \in \mathbb{N}$, the amplitude Eq.~\eqref{eq:Etot2} becomes proportional to the quality factor $Q$, as expected. Additionally, as it is also proportional to the injected amplitude $X_a$, we directly see the interesting feature of looking at the field squared amplitude, the applied field acting as an amplification for the DM field.

Finally, the signal we look for is a slow oscillation in the transition frequency of Rydberg atoms with respect to an unperturbed reference 
\begin{equation}\label{signal_stark}
    \nu(t) = \nu_0 + \Delta\nu \cos\left(\Delta \omega t +\phi\right) \, ,
\end{equation}
whose amplitude is given by
\begin{equation}\label{signal}
\Delta \nu = - \chi \frac{\Delta\alpha S(\omega,\omega_a;\rho_\mathrm{DM},  X_a;L,r)}{2h} \, ,
\end{equation}
which depends on both the applied external field and the DP field.

\section{\label{Experimental_method}Experimental considerations}

\subsection{Rydberg atoms}\label{sec:Rydberg}

The transition frequency measurement can be performed using a regular atomic clock, which allows very good uncertainty on the frequency measurement, but is not very sensitive to the Stark effect e.g. for the $5s^2\,^1S_0 \to 5s5p\,\,^3P_0$ clock transition in Sr the differential polarizability $\Delta \alpha/2h \approx 3.1\times 10^{-6}$~Hz/(V/m)$^2$ \cite{Middelmann}, thus requiring a strong applied field.

To overcome this lack of sensitivity to $E^2$, ``regular'' atoms can be replaced by Rydberg atoms, which are in a quantum state with high principal quantum number $n$ \cite{Gallagher_1988}. The electrons are much further from the nucleus thus the atom has much higher polarizability. For large $n$ the corresponding polarizabilities in Sr can reach $\Delta\alpha/2h \approx 10^{5}$~Hz/(V/m)$^2$ ($n>60$) \cite{Millen2011}.

In \cite{Bridge2016} the measurement of the transition probability for a given frequency lasts about 300 $\mu$s. At least three such measurements are necessary to fully determine the resonance (amplitude, width and center frequency), which implies a maximum sampling rate of about 1~kHz, which is the maximum value we will assume. In principle the process could be faster with higher laser power and/or nondestructive techniques. 

Typically, Sr Rydberg measurements use laser cooled Sr atoms that are excited to Rydberg states using two photon transitions \cite{Millen2011,Bridge2016} and direct spectroscopy of e.g. the $5s5p\,^3P_1 \to 5sns\,^3S_1$ transition is performed, with $n$ up to 81 \cite{Millen2011,Bridge2016}. In regular dispersive measurements, the photon scattering rate per atom, i.e the rate at which an atom absorbs and re-emits incident photons, is high implying that after a single detection, the atom is too hot,
and no longer trapped for a second detection. In that situation, the measurement is said to be destructive. An alternative method is a nondestructive measurement, based on a low photon scattering rate, meaning that a single atom can be used for multiple measurement. This non-destructive process has already been experimentally tested, and is based on a differential dispersive measurement \cite{Vallet}. As a consequence, it is not necessary to produce new Rydberg atoms for each frequency measurement, implying that the 1~kHz sampling frequency is feasible. 

\subsection{\label{bandwidth_exp}Bandwidth of detection}

In order to see the oscillations at $\Delta\omega$ in the total field power Eq.~\eqref{general_E_power}, we require the DM and the applied fields to have different frequencies. We also require the Nyquist  frequency of the apparatus to be higher than the angular frequency of the signal $\Delta \omega < \pi f_s$ to be able to detect any oscillatory behavior in the transition frequency of the atoms.

For a given applied angular frequency $\omega_a$, sampling frequency $f_s$, Rydberg atoms perform the measurements of the electric field squared during $\Tobs$ at an angular frequency $\Delta \omega$ (more precisely, $\Tobs\times f_s$ measurements will be taken for each $\omega_a$). The time $\Tobs$ is arbitrary; if it is longer than the measurement process comprising excitation and ionisation, one has to prepare again the atoms to their Rydberg state after deexcitation accordingly. As detailed above, the experiment is sensitive to any $\Delta \omega$ such that $2\pi/\Tobs \leq |\Delta \omega| \leq \pi f_s$, which, in terms of DM angular frequency, is equivalent to
\begin{align}
\omega  \in [\omega_a - \pi f_s;\omega_a - \frac{2\pi}{\Tobs}] \cup [\omega_a + \frac{2\pi}{\Tobs};\omega_a + \pi f_s] \, .
\end{align}
At the end of the individual measurement time $\Tobs$, the second step would be to shift the applied angular frequency $\omega_a$ by the sampling angular frequency, $2\pi f_s$ and make a second measurement of the electric field squared for DM angular frequencies $\omega$ around this new $\omega_a$, during $\Tobs$ again. This way, except at the exact angular frequencies of the applied field $\omega_a$, all possible DM frequencies are scanned. 

This scheme can be repeated N times, as much as time allows. At the end of this loop, the total experimental time is simply $T_{\mathrm{tot}} = N\Tobs$. The corresponding total DM frequencies band scanned is $f_{\mathrm{tot}} = Nf_s = T_{\mathrm{tot}}f_s/\Tobs$. The larger the total experimental time, the larger the band of scanned DM frequencies. Additionally, the blind spots at exact $\omega_a$ can be avoided, and sensitivity can be optimized (see below) by shifting $\omega_a$ by a little less than $2\pi f_s$, at the expense of increasing the overall experimental duration. 

The coherence time of the DM field is inversely proportional to its Compton frequency and follows \cite{Derevianko2018}
\begin{align}
    \tau(\omega) = \frac{c^2}{\sigma^2_v \omega}\approx \frac{10^6}{\omega} \, ,
\end{align}
where we considered the usual DM velocity distribution with dispersion $\sigma_v \sim 10^{-3}c$. For DM frequencies in the GHz range, this coherence time is much smaller than $T_{\mathrm{obs}}$, which needs to be taken into account in the sensitivity estimate (see Eq.~\eqref{eq:Emin}).

\section{\label{Noises}Sensitivity estimate}
In this section, we will present the level to which a realistic experiment can detect or constrain the coupling parameter $\chi$. First, we will focus on the measurement noise and present some limits on $\chi$ reachable considering only statistical noise. We will then discuss the main systematics related to this experiment: the intensity variations of the applied field. Finally, by using realistic values for the cavity, laser and various noise sources, we will estimate the detection limit on $\chi$ reachable for a dedicated experiment.

\subsection{Statistical measurement noise} \label{sec:stat-noise}
The first source of noise limiting the sensitivity of the experiment is the statistical noise related to the measurement of the frequency shift experienced by Rydberg atoms under the perturbation from an external electric field. In \cite{Bridge2016} the reported resolution of the spectroscopy of $n=56$ Rydberg states in Sr is of the order of a few kHz,  at a maximum possible sampling rate of 1~kHz (see Sec. ~\ref{sec:Rydberg}).

We will thus consider two scenarios for our order of magnitude estimates of the experimental sensitivity. One with a ``modest'' sampling rate of 100~Hz and a second, more optimistic one, with higher sampling at $f_s=1$~kHz. In both cases we will assume a single shot spectroscopic resolution of $\sim$1~kHz for differential polarizabilities of $\Delta\alpha/2h \approx 10^5$~Hz/(V/m)$^2$, corresponding to Rydberg states with principle quantum numbers $n\sim 60-70$ \cite{Millen2011}.

We denote the power spectral density (PSD) measurement noise of $E^2$ as $S_{E^2}$, which translates into a minimal detectable power of the total field inside the cavity of 
\begin{align}\label{eq:Emin}
     E^2_\mathrm{min} = \sqrt{\frac{2S_{E^2}}{\sqrt{\Tobs \tau(\omega)}}} \, ,
\end{align}
where $\Tobs$ is the individual integration time and $\tau(\omega)$ is the coherence time of the field such that $\Tobs \gg \tau(\omega)$, as discussed in the previous section.

This means, that for a signal-to-noise ratio (SNR) of 1, the constraint on $\chi$ for a mass of the DP $m$ corresponding to an angular frequency $\omega$ can be computed by equating Eqs.~(\ref{eq:Etot2}) and \eqref{eq:Emin}, which leads to 
\begin{align}
\left[\chi(\omega)\right]_\mathrm{stat} =& \frac{E^2_\mathrm{min}} {S(\omega,\omega_a;\rho_\mathrm{DM}, X_a;L,r)} \,
 \nonumber \\
=& \frac{\sqrt{2S_{E^2}}}{(T_\mathrm{obs}\tau(\omega))^{\frac{1}{4}}S(\omega,\omega_a;\rho_\mathrm{DM}, X_a;L,r)}\, . 
\label{eq:chi_stat}
\end{align}

\subsection{\label{Systematic} Amplitude fluctuation of the applied field}

The main systematic identified for this experiment comes from fluctuations of the amplitude of the applied electromagnetic field.  Indeed, the principle of the experiment consists in measuring oscillations of the electric field intensity at the center of the cavity. Amplitude fluctuations of the injected electromagnetic field will mimic such a signal and can jeopardize the results of the experiment. In this section we assume that the main source of fluctuations of the field inside the cavity are fluctuations of the power of the signal that is fed into the cavity, i.e. relative intensity noise (RIN) of the signal generator.

We model the amplitude of the injected electric field by including a stochastic component, i.e. replacing the previously considered constant $\vec X_a$ by
\begin{align}
    \vec X_a \rightarrow \vec X_a\left[1+\int d\omega_0 \frac{\Delta X_a(\omega_0)}{X_a}\cos(\omega_0 t + \phi_0)\right] \, .
    \label{stochastic_noise}
\end{align}

In this expression, $\Delta X_a\left(\omega_0\right)$ is a stochastic contribution modeling the spectral amplitude of the noise characterized by the RIN PSD denoted $S_\mathrm{RIN}(\omega)$ and defined by
\begin{equation}\label{eq:PSD_noise}
\frac{\Delta X_a(\omega_0)}{X_a} \equiv \sqrt{\frac{S_\mathrm{RIN}(\omega_0)}{2T_\mathrm{obs}}} \, .
\end{equation}
where the factor 2 in the denominator arises because the RIN is a fluctuation of the laser power, not its amplitude.
Typically, the RIN of frequency generators in the microwave domain (GHz frequencies of interest here) is characterized by a flicker noise, see e.g. \cite{Rubiola}, such that we can parametrize its PSD as 
\begin{align}
    S_\mathrm{RIN}(\omega) = \frac{P_\mathrm{RIN}}{\omega} \, ,
    \label{eq:flicker_noise}
\end{align}
where $P_\mathrm{RIN}$ is dimensionless.

Let us now show how the $\Delta X_a$ fluctuations can produce an harmonic signal in $\left|\vec E\right|^2$ of angular frequency $\Delta \omega=\omega_a - \omega$, i.e. mimick the searched signal of Eq.~(\ref{signal_stark}). 

We will work to leading order in $\Delta X_a/X_a$ and in particular neglect terms that are $\mathcal O\left(\left(\frac{\Delta X_a}{X_a}\right)^2\right)$ and $\mathcal O\left(\frac{\Delta X_a}{X_a}\frac{X_\mathrm{DM}}{X_a}\right)$. Considering the modification of applied field amplitude Eq.~\eqref{stochastic_noise}, the fluctuation $\Delta X_a$ will only be considered at frequencies $\omega_0$ producing a noise in the electric field squared at angular frequency $\Delta \omega$. These angular frequencies are $\omega_0\simeq \{\Delta \omega; 2\omega_a\}$. Considering the RIN as a flicker noise characterized by a PSD of the form Eq.~\eqref{eq:flicker_noise}, the fluctuation amplitude at $\omega_0=\Delta \omega$ will be multiple order of magnitudes larger than its amplitude at $\omega_0=2\omega_a$. For this reason, in the following, we will only consider the fluctuation at frequency $|\Delta X_a(\omega_0=\Delta \omega)|$ such that the amplitude of the applied field Eq.~\eqref{stochastic_noise} becomes 
\begin{align}
    \left(\vec X_a+\Delta \vec X_a(\Delta \omega)\cos(\Delta \omega t+\phi_0)\right)\cos(\omega_at + \phi) \, .
    \label{stochastic_noise_single}
\end{align}

We now use the same procedures as the one described in Sec~\ref{sec:model_E} to compute the RIN contribution to the total electric field at the center of the cavity. This calculation is presented in details in Appendix~\ref{ap:applied_field} and leads to 
\begin{align}
\left[\vec E^\mathrm{tot}_a\right]_\mathrm{RIN} &= \frac{\Delta X_a(\Delta \omega)}{2X_a}\left( \vec A(2\omega_a-\omega)\cos([2\omega_a-\omega]t+\phi_+) \nonumber \right.\\
& \left. +\vec B(2\omega_a-\omega)\sin([2\omega_a-\omega]t+\phi_+) \right. \, \\
&\left.+\vec A(\omega)\cos(\omega t+\phi_-) +\vec B(\omega)\sin(\omega t+\phi_-)\right) \, ,\nonumber
\end{align}
in addition to the contribution from Eq.~\eqref{eq:E_tot_a} and where $\phi_\pm=\phi \pm\phi_0$. 

Then, the RIN contribution to the total field power at the center of the cavity can be obtained by multiplying the last equation with Eq.~(\ref{eq:E_tot_a}). Keeping only the terms oscillating at angular frequency $\Delta \omega$, the RIN contribution to $E^2$ is given by
\begin{widetext}
\begin{subequations}
\begin{align}
\left[ E^2(\omega,\omega_a)\right]_\mathrm{RIN}&=\frac{\Delta X_a(\Delta \omega)}{2X_a}\left(\left(A(\omega_a)\left[A(2\omega_a-\omega)+A(\omega)\right]+B(\omega_a)\left[B(2\omega_a-\omega)+B(\omega)\right]\right)\cos(\Delta \omega t+\phi_0)\,\right.\nonumber \\
&\left.\qquad\quad + \left(A(\omega_a)\left[B(2\omega_a-\omega)-B(\omega)\right]+B(\omega_a)\left[A(\omega)-A(2\omega_a-\omega)\right]\right)\sin(\Delta \omega t+\phi_0)\right)  \label{full_E_power_noise}\\
& = \frac{\Delta X_a (\Delta \omega)}{2X_a}\sqrt{N(\omega,\omega_a)} \cos(\Delta \omega t + \varphi)= \sqrt{\frac{P_\mathrm{RIN} N(\omega,\omega_a)}{8 T_\mathrm{obs} \Delta \omega}} \cos(\Delta \omega t + \varphi)\, ,
\label{general_E_power_noise}
\end{align}
\end{subequations}
\end{widetext}
where we have used Eqs.~(\ref{eq:PSD_noise}) and (\ref{eq:flicker_noise}). In this expression, the functions $A$ and $B$ are the norm of $\vec A$ and $\vec B$ whose complete expressions are given by Eqs.~(\ref{eq:A_B}). The function $\sqrt{N(\omega,\omega_a)}$ correspond to the noise amplification factor by the cavity (see Appendix \ref{ap:signal_noise_amp_annex}), quadratic in $X_a$ and whose expression is given by Eq.~(\ref{noise_amp})  and $\varphi$ is a phase. 

The RIN contribution to $E^2$ from Eq.~(\ref{general_E_power_noise}) will limit the sensitivity of the experiment to values of $\chi$ that makes the signal from Eq.~(\ref{eq:Etot2}) larger than the systematic (i.e. larger than Eq.~(\ref{general_E_power_noise}). In other words, the RIN will limit the sensitivity of the experiment to value of $\chi$ that are larger than
\begin{equation}\label{eq:chi_RIN}
    \left[\chi(\omega)\right]_\mathrm{RIN} = \frac{\sqrt{\frac{P_\mathrm{RIN} N(\omega,\omega_a)}{8 T_\mathrm{obs} \Delta \omega}}}{S(\omega,\omega_a;\rho_\mathrm{DM}, X_a;L,r)} \, ,
\end{equation}
where the function $S$ is defined in Eq.~(\ref{eq:Etot2}). Note that this limit is linear in $X_a$, the amplitude of the applied electric field.

Note that vacuum fluctuations noise, i.e shot noise, can also be viewed as an amplitude fluctuation of the electric field squared, as seen by Rydberg atoms, thus is analogous to the RIN. However, it is negligible since its PSD normalized by the number of photons $N$ $\sqrt{S_\mathrm{SN}}/N \sim 10^{-11} \: (\mathrm{/\sqrt{Hz})} \ll \sqrt{S_\mathrm{RIN}}$ (using experimental parameters considered in Sec.~\ref{sensitivity} and a typical mode radius of $\sim 0.01$ m).

\subsection{\label{sec:opti_param}Optimum choice of experimental parameters}

The sensitivity of the experiment at a given angular frequency $\omega$ relies on the signal amplitude Eq.~\eqref{eq:Etot2} but also on the limiting noise. Combining Eq.~\eqref{eq:chi_stat} and Eq.~\eqref{eq:chi_RIN}, this is simply
\begin{align}
\left[\chi(\omega)\right]_\mathrm{limit} &= \frac{\sqrt{\frac{2S_{E^2}}{\sqrt{\tau(\omega)}}+\frac{P_{\RIN}N(\omega,\omega_a)}{8\Delta \omega \sqrt{T_\mathrm{obs}}}}}{\Tobs^{\frac{1}{4}}S(\omega,\omega_a;\rho_\mathrm{DM}, X_a;L,r)} \,,
    \label{eq:chi_limit}
\end{align}
where we used the usual quadratic sum of uncertainties since the two contributions are uncorrelated.

The maximum angular frequency difference $\Delta \omega$ corresponds to, from Sec.~\ref{bandwidth_exp}, half the sampling frequency ($=\pi f_s$). To better understand the sensitivity of the experiment, we will simplify the expressions of $S(\omega,\omega_a;\rho_\mathrm{DM}, X_a;L,r)$ and $N(\omega,\omega_a)$ in Eqs.~\eqref{eq:Etot2}, ~\eqref{general_E_power_noise} and \eqref{noise_amp}, considering $r = 1 - \epsilon$, $\epsilon \ll 1$ and $\omega \sim \omega_a$\footnote{Even in the case where $\Delta \omega = \pi f_s$, we still consider $\omega \sim \omega_a$ in this approximation, as $f_s \lessapprox \frac{\omega}{Q} \approx \omega(1-r) \ll \omega$. For a sampling frequency $\sim$~kHz and Compton frequency in the GHz domain, and $r = 1 -10^{-6} \sim 1$}. In that case, 
\begin{subequations}
\begin{align}
    N(\omega_a) &\approx \frac{X^4_a\epsilon^2}{\cos^4(\frac{\omega_a L}{2c})}\, , \label{syst_noise_approx}\\
    S(\omega_a) &\approx \frac{\beta c X_a \sqrt{2\mu_0 \rho_\mathrm{DM}}\sqrt{\epsilon}}{\sqrt{2}\cos^2(\frac{\omega_a L}{2c})}\left(1+\cos\left(\frac{\omega_a L}{2c}\right)\right)\,,\label{signal_approx}
\end{align}
\end{subequations}
at lowest order in $\epsilon$.
We see from \eqref{syst_noise_approx} that the two uncertainties in the numerator of \eqref{eq:chi_limit}) depend differently on $X_a$ (the statistical uncertainty is independent of $X_a$, the RIN contribution is quadratic in $X_a$), while the signal strength in the denominator is linear in $X_a$. This suggests an 'optimum' value of $X_a$ such that Eq.~\eqref{eq:chi_limit} is minimum,
\begin{subequations}
\begin{align}
    \frac{d\chi(\omega_a)}{dX_a} &= 0\,\\
   \Rightarrow X_a(\omega_a) &\approx \sqrt[4]{\frac{16S_{E^2} \Delta\omega}{ \epsilon^2 P_{\RIN}}\sqrt{\frac{T_\mathrm{obs}}{\tau(\omega_a)}}}\left|\cos\left(\frac{\omega_a L}{2c}\right)\right| \,\\
   &\approx \sqrt[4]{\frac{16\pi f_s S_{E^2}\sqrt{\omega_a T_\mathrm{obs}}}{10^3 \epsilon^2 P_{\RIN}}}\left|\cos\left(\frac{\omega_a L}{2c}\right)\right|\, ,\label{optimum_amplitude}
\end{align}
\end{subequations}
where we used Eqs.\eqref{syst_noise_approx},\eqref{signal_approx}. We considered the maximum angular frequency shift between the DM and applied frequencies $\Delta \omega =\pi f_s$, in order to have the best sensitivity on $\chi(\omega)$. Note that this equation to approximate the optimum value of $X_a$ is not valid for angular frequencies $\omega_a L/c = 2\pi + 4n\pi, n \in \mathbb{N}$ and $\omega_a L/c = \pi + 2n\pi, n \in \mathbb{N}$. In the first case, the signal decreases significantly (see Eq.~\eqref{signal_approx}) and the experiment becomes insensitive to DM, while in the second case, Eq.~\eqref{optimum_amplitude} would indicate to apply $X_a=0$, which would automatically set the signal to 0, at first order in $\chi$. 

From Eq.~\eqref{optimum_amplitude}, and still considering $\omega \sim \omega_a$, we can express the sensitivity of the experiment $\chi(\omega)$ as
\begin{align}
\chi(\omega) &\approx \frac{\sqrt{2}\left|\cos\left(\frac{\omega L}{2c}\right)\right|}{1+\cos\left(\frac{\omega L}{2c}\right)}\left(\frac{P_{\RIN}S_{E^2}}{10^3\pi f_s }\right)^{\frac{1}{4}}\frac{\left(\omega\Tobs^{-3}\right)^{\frac{1}{8}}}{\beta c\sqrt{2\mu_0 \rho_\mathrm{DM}}}\, ,
\label{optimum_chi}
\end{align}
for $\frac{\omega L}{c} \neq \pi + 2\pi n$, $n \in \mathbb{N}$. 
This is a simplified expression that provides an approximate evaluation of the optimal sensitivity of the experiment.

In the following section, we will assume that the applied field amplitude $X_a$ is modified each time its angular frequency $\omega_a$ is shifted, so that the condition Eq.~\eqref{optimum_amplitude} is always fulfilled, but we use the full expressions of $S(\omega,\omega_a;\rho_\mathrm{DM}, X_a;L,r)$ and $N(\omega,\omega_a)$ in Eqs.~\eqref{eq:Etot2} and \eqref{general_E_power_noise} to evaluate the sensitivity.

\subsection{\label{sensitivity}Reachable sensitivity in a realistic experiment}

\begin{figure*}[ht!]
    \centering
    \captionsetup{justification=centering}
    \includegraphics[width=0.9\textwidth]{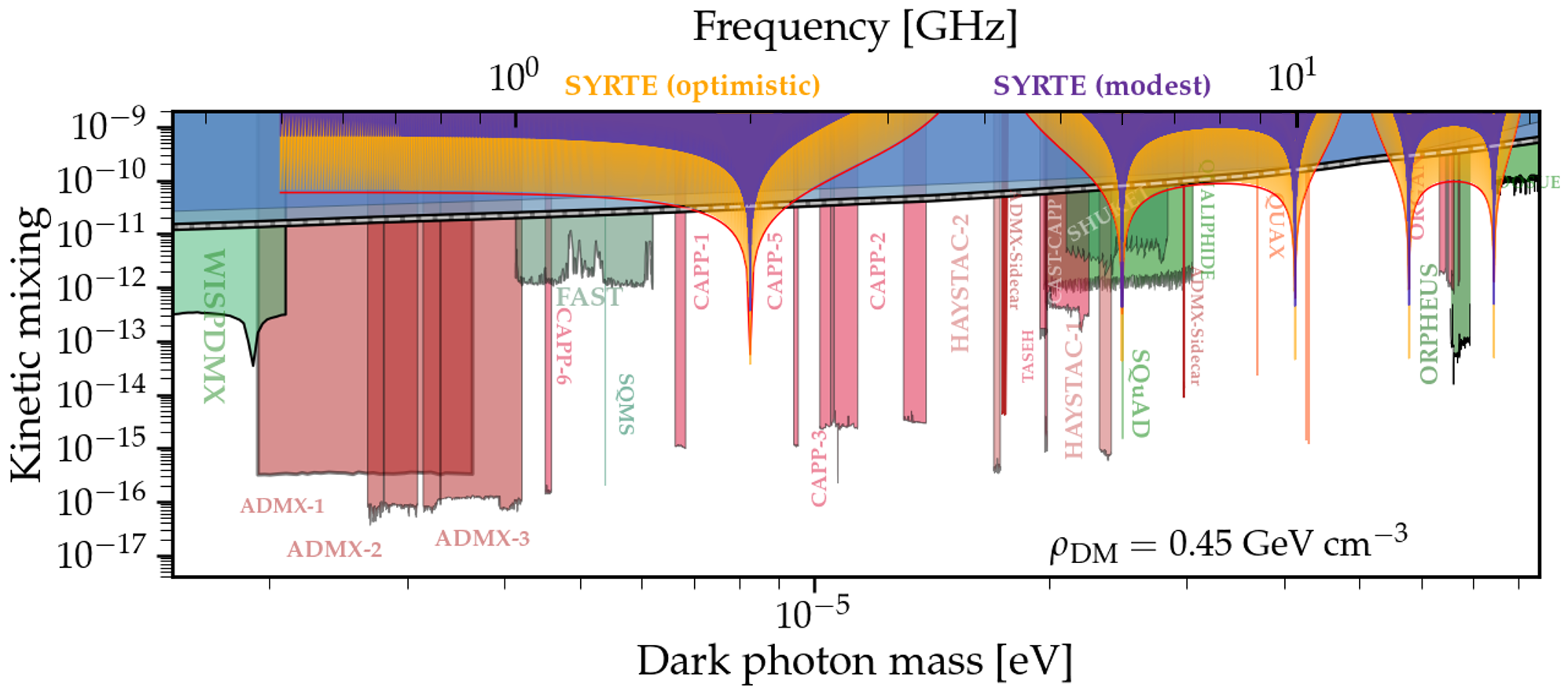}
    \caption{\label{fig:constraints}Current constraints on dark photons through their kinetic mixing coupling $\chi$ with photons (renormalized from \cite{AxionLimits,Caputo}, see text). The expected sensitivity of this work, with "optimistic" parameters of the system described in Table ~\ref{Table_microwave} is shown in orange, denoted "SYRTE (optimistic)", while the sensitivity with the "modest" parameters of Table ~\ref{Table_microwave} is shown in purple, denoted "SYRTE (modest)" (see text). The bordering red line around the orange curve indicates the approximate sensitivity of the experiment in the optimistic case, computed from Eq.~\eqref{optimum_chi}.}
\end{figure*}

We now estimate the sensitivity on $\chi$ of the experimental method defined in Sec.~ \ref{Experimental_method} by considering the two noise sources described  in Sec.\ref{Noises}. The measurement noise will lead to a  lower limit on $\chi$ given by Eq.~(\ref{eq:chi_stat}) while the main systematic effect, the RIN, will lead to a lower limit on $\chi$ given by Eq.~(\ref{eq:chi_RIN}).

First, we consider the random polarized case for the DP polarization direction, i.e $\beta=1/\sqrt{3}$. Some already existing constraints come from experiments with this assumption \cite{SHUKET,SquAD,WISPDMX,Tokyo1,Tokyo2,FUNK,Qualiphide,FAST}. Other experiments \cite{ADMX:2010,ADMX:2018_1,ADMX:2019, ADMX:2021, ADMX_Sidecar:2018,ADMX_Sidecar:2021,Alesini:2019,Alesini:2020,Alesini:2022,Capp:2020_1,Capp:2020_2,CAPP:2020_3,Capp:2022_1,Capp:2022_2, Capp:2022_3} considered the fixed polarization scenario, and we have rescaled them accordingly (using \cite{Caputo}). Finally, some of them assumed perfectly aligned polarizations between DP and device, so they also have been rescaled \cite{Tokyo1,DM_PathFinder}\footnote{For ORGAN \cite{Organ} and its $\mathcal{O}$(month) of data taking, this correction factor is of $\mathcal{O}$(1), as for long time experiment, the sensitivity in both scenarios is equal (Sec. VI.6 of \cite{Caputo}).}

We consider the local DM energy density to be $\rho_\mathrm{DM} =$ $0.45 \mathrm{\ GeV/cm}^3$ \cite{Caputo}.

Let us remind that the observational scheme considered here consists in electric signals of angular frequency $\omega_a$. For each injected electric field we perform a measurement of duration $T_\mathrm{obs}$ and then shift the angular frequency of the applied field by $\pi f_s$. Each measurement of duration $T_\mathrm{obs}$ provides constraints in the frequency range $\{\omega_a-\pi f_s; \omega_a+\pi f_s\}$ in steps of $2\pi/T_\mathrm{obs}$.

Let us now consider some numbers for the experiment, which are all summarized in Table \ref{Table_microwave}. First of all, we consider an individual measurement duration of $\Tobs= 60$ s. This duration is arbitrary, but we list two important considerations for this choice. We require $\Tobs$ to be large for best sensitivity (see Eq.~\eqref{eq:chi_limit}), but short enough to allow scanning a large range of DM frequencies in a reasonable amount of time ($\leq \mathcal{O}$(month)).

As described in Section \ref{sec:stat-noise}, we consider two different sampling rates which lead to different measurement noise PSD and directly affect the sensitivity of the experiment (cf Eq.~\eqref{optimum_chi}).
To put some numbers, assuming a single shot spectroscopic resolution of $\sim$1~kHz \cite{Bridge2016} for differential polarizabilities of $\Delta\alpha/2h \approx 10^5$~Hz/(V/m)$^2$, leads to a noise PSD of the measured electric field power of $S_{E^2}\approx 10^{-6}$~(V/m)$^4$/Hz for $f_s=100$~Hz and $S_{E^2}\approx 10^{-7}$~(V/m)$^4$/Hz for $f_s=1$~kHz. 
Regarding the systematic effect, the amplitude of the flicker noise can be considered to be $P_{\RIN} = 10^{-13}$ (based on ``off the shelf'' components studied in \cite{Rubiola} more than a decade ago) in the modest case, and we assume an improved RIN control for the optimistic case, with an amplitude of $P_{\RIN} = 10^{-15}$. The level of the systematic effect PSD is also impacted by the sampling rate, following Eq.~\eqref{general_E_power_noise}. Moreover, as derived in Sec.~\ref{sec:opti_param}, the approximate sensitivity of the experiment Eq.~\eqref{optimum_chi} scales as $(P_\mathrm{RIN}S_{E^2}/f_s)^{1/4}$, implying that the modest and optimistic scenarios will differ in the sensitivity by a factor $\sim$10. 
The optimum value of the amplitude of the applied field $X_a$ is then derived from Eqs.~ \eqref{eq:chi_limit} (for the minimum value\footnote{The smallest value of $X_a$ corresponds to DM/applied frequencies close to odd modes, from Eq.~\eqref{optimum_amplitude}. In this regime, both approximate amplitudes of noise Eq.~\eqref{syst_noise_approx} and signal Eq.~\eqref{signal_approx} reach infinity, implying we must consider the real expression of $\chi(\omega)$ derived in Eq.~\eqref{eq:chi_limit} and check for which value of $X_a$ the sensitivity on $\chi$ is the highest.}) and \eqref{optimum_amplitude} and all other experimental parameters. Since it depends on the DM Compton frequency, we provide the range of optimal $X_a$, for the modest case
\begin{align}
    18.1 \,\,\mathrm{V/m} &\lesssim X_a \lesssim 1.70 \times 10^5 \ \mathrm{V/m} \, .
\end{align}
It is independent of the sampling frequency since, from Eq.~\eqref{optimum_amplitude}, $X_a(\omega_a) \propto (f_s S_{E^2})^{1/4}$ and $S_{E^2}\propto f^{-1}_s$, but not of the systematic effect level $P_\mathrm{RIN}$.
\begin{table}[ht!]
\caption{Assumed experimental parameters}
\centering
\begin{tabular}{c c}
\hline \hline
Parameters & Numerical values \\
\hline \hline
Quality factor $Q$ \cite{SYRTE_fountains} & $10^4$ \\
Mirrors reflectivity r & $\approx 1-2 \times 10^{-4}$ \\
Cavity length L & $7.5 \mathrm{\ cm}$\\
Injected field strength $X_a(\omega)$ &  $[18.1,1.70\times 10^5]$ V/m \\
Sampling frequency $f_s$ & $10^2 \, ; \, 10^3$ Hz\\
Individual measurement time $T_\mathrm{obs}$ & 60 s \\
Range of $f_a=\omega_a/2\pi$ & $[0.5,20.5]$ GHz\\
Range of $\Delta \omega$ & $[2\pi/\Tobs,\pi f_s]$ rad/s \\
Statistical noise PSD $S_{E^2}$ & $ 10^{-4}/f_s \,\, (\mathrm{V/m})^4/\mathrm{Hz}$ \\
Systematic effect PSD $S_\mathrm{RIN}(\omega)$ & $10^{-13}/\omega \,; \, 10^{-15}/\omega$ \\
\hline
\end{tabular}
\label{Table_microwave}
\end{table}

The sensitivity of the experiment obtained considering all experimental parameters described in Table~(\ref{Table_microwave}) and respectively with $\{f_s=1$ kHz, $P_\mathrm{RIN}=10^{-15}\}$ and $\{f_s=100$ Hz, $P_\mathrm{RIN}=10^{-13}\}$ is presented by the orange and purple curves of Fig.~\ref{fig:constraints}. One can clearly see the sensitivity peaks arising from the cavity's odd resonances. This happens when the applied field amplitude $X_a$ is small, as shown in Eq.~\eqref{optimum_amplitude}. This equation works well for frequencies far from odd resonances. However, on those odd resonances, this approximate equation cannot be used as discussed previously. Instead one should use the exact expressions of signal and noise to optimize Eqs.~\eqref{eq:chi_limit}. As an example, when the applied field frequency corresponds exactly to the first odd resonance of the cavity, i.e $\omega_a L =\pi c$ and $\omega = \omega_a + \pi f_s$ the optimum amplitude of $X_a$ is $\sim 18.1$ V/m, whose corresponding experimental sensitivity is $\chi \sim 10^{-13}$ in the modest scenario, as shown in Fig.~\ref{fig:constraints}.
Additionally, one can notice the presence of specific frequencies where this sensitivity decreases significantly, the experiment is almost insensitive to these DM frequencies. As discussed in the previous section, from the approximate expression of the signal contribution Eq.~\eqref{signal_approx}, we have $S(\omega,X_a) \simeq 0$ for $\frac{\omega L}{c} = 4\pi + 2 \pi n $, $n \in \mathbb{N}$ accounting for the loss of sensitivity.  
In both scenarios presented here (modest and optimistic), one can see from Fig.~\ref{fig:constraints} that the experiment setup proposed here would improve the current constraint on the coupling $\chi$ compared to cosmological and astrophysical observations and other already existing laboratory experiments.

\section{Discussion and conclusion}

If one decides to run this experiment aiming at unconstrained regions of the exclusion plot, it would take approximately five days of data-taking to cover the mass range from 7 $\mu$eV to 10 $\mu$eV, while around 35 days would be needed to cover the mass range from 35 $\mu$eV to 60 $\mu$eV, assuming no dead time between the $T_\mathrm{obs}=60$~s observation runs. More realistically, reserving say 50\% of the total experimental time for manipulation of the atoms and applied field, the total duration increases by a factor two, which is still very reasonable.

With the appropriate set of parameters, in particular the applied field amplitude $X_a$ following Eq.~\eqref{optimum_amplitude}, both sources of noise, systematic and statistical, are equal in amplitude. This means that, in the search for high sensitivity of the experiment, the optimum choice of $X_a$ is not to increase it as much as possible to maximize the signal. Even though the signal is linear in $X_a$, the systematic uncertainty is quadratic in $X_a$, as stated at the end of Sec.~\ref{Systematic}, implying a loss of sensitivity if the experimenter decides to apply too much power inside the cavity.

If the level of intensity fluctuations (RIN) of the applied field could be reduced e.g. by stabilizing the power using low noise intensity measurements \cite{Orpheus}, the applied field and/or the quality factor of the cavity could be increased leading to an increase of the signal whilst keeping the contribution from the RIN below that of the measurement noise in Eq.~\eqref{eq:chi_limit}. This way, the optimistic curve presented in Fig.~\ref{fig:constraints} would be achievable.

Some experiments use curved mirrors (\cite{SHUKET}, \cite{Tokyo1}, \cite{Tokyo2}) to focus the DM induced electric field into a reduced region, to be able to detect more power. This method is not considered here, but with the appropriate curvature, it may improve the sensitivity of the experiment.

In conclusion, we propose a novel experiment that uses Rydberg atoms inside a microwave cavity to search for DP through its mixing with standard electromagnetism. Our proposal features an optimized applied electric field inside the cavity and Rydberg atoms as highly sensitive probes of $E^2$. This latter feature allows searching for the cross-term between the applied and DP-induced fields, thus allowing best sensitivity with relatively low sampling frequencies.

Using realistic experimental parameters we show (see Fig.~\ref{fig:constraints}) that such an experiment has the potential to significantly improve on existing laboratory experiments in terms of the sensitivity, and more importantly, in terms of the explored DP-mass regions. The latter can be more specifically targeted by tuning the cavity size such that resonances match the least explored regions. Finally, we note that around the resonances the experimental sensitivity is also better than indirect bounds coming from cosmological considerations (blue region in Fig.~\ref{fig:constraints}).   

\begin{acknowledgments}
The authors acknowledge S\'ebastien Bize, Luca Lorini and Giorgio Santarelli for very helpful discussions. This work was supported by the Programme National GRAM of CNRS/INSU with INP and IN2P3 cofunded by CNES.
\end{acknowledgments}

\appendix

\section{\label{app:static_DP}Does the DP field propagate or not?}

Let us assume the dark photon vector field is an oscillating field which does not propagate.  In this case, 
\begin{equation}
\vec{E}_\mathrm{DM} = \vec{X}_\mathrm{DM}\cos(\omega t) \, .
\end{equation} 
This statement is true in the rest frame of dark matter with constant mode k = 0. If we consider it moving with respect to Earth at a velocity $v=10^{-3}$ c, which is the typical mean velocity of the dark matter halo, we have 
\begin{equation}
\vec{E}_\mathrm{DM} = \vec{X}_\mathrm{DM}\cos(\omega' t - \vec{k'}.\vec{x}) \, ,
\end{equation} 
and from special relativity, we have
\begin{subequations}
\begin{align}\label{SR_transf}
\omega' &= \gamma(\omega - \vec{v}.\vec{k}) \approx \omega  \, , \\
\vec{k}' &= \vec{k} + \frac{1}{v^2}(\gamma -1)(\vec{v}.\vec{k})\vec{v}-\frac{1}{c^2}\gamma \omega \vec{v} \approx - \frac{1}{c^2} \omega \vec{v} \, , 
\end{align}
\end{subequations}
since $\gamma=1-\frac{v^2}{2c^2} \sim 1$. To neglect this propagation term, we require $2\pi \gg k'L$, with $L$ the length of the cavity, or equivalently, the de Broglie wavelength of the field has to be much bigger than the size of the cavity. Considering a length $\mathcal{O}$(10 cm), the corresponding constraint on $k'$ is
\begin{align}
k' \ll 20\pi \mathrm{\ m}^{-1} \, .
\end{align}
In this paper,  we are interested in DP masses in microwave domain, $10^{-6} < mc^2 \mathrm{\ (eV)}< 10^{-4}$, or
\begin{align}
k' \in [0.05, 5] \times 10^{-1} \mathrm{\ m}^{-1}  \, ,
\end{align}
implying we can safely neglect the propagation term.
Note that if we were to work in the optical frequency range, this propagation term should be kept, as in this case, $k' \in [5, 25] \times 10^6 \mathrm{\ m}^{-1} \gg 2\pi/L$.

\section{\label{Fields_expression}Full calculation of field amplitude in the cavity}

\subsection{Applied field}\label{ap:applied_field}

The experimental setup (Fig. \ref{DP_cavity}) gathers an applied field $E_a$ with amplitude $X_a$, angular frequency $\omega_a$ and a time varying amplitude fluctuation $\Delta X_a$ with angular frequency $\omega_0$, considered much smaller than $\omega_a$. This field has random phase $\phi$ with respect to the DM induced oscillating field $\vec E_{\mathrm{DM}}$. Considering application of this field at the left edge of the cavity, and assuming a transmission coefficient of the mirror being $t=\sqrt{1-r^2}$, the first contribution at the center of the cavity reads
\begin{widetext}
\begin{align}
\vec E^0_a(x=0,t) &= \Re \left[t\left(\vec{X}_a e^{-i(\omega_at-k_a\frac{L}{2}+\phi)}+\frac{\Delta \vec{X}_a}{2}\left(e^{-i(\omega_+t-k_+\frac{L}{2}+\phi_+)}+e^{-i(\omega_-t-k_-\frac{L}{2}+\phi_-)}\right)\right)\right]\, ,
\end{align}
\end{widetext}
where $k_\pm = k_a \pm k_0$. This contribution propagates until the other cavity boundaries, gets reflected once with coefficient $-r$ such that boundary conditions are respected, then comes back to the center, implying that the second contribution reads 
\begin{widetext}
\begin{align}
\vec E^1_a(x=0,t) &= \Re\left[-tre^{ik_aL}\left(\vec{X}_a e^{-i(\omega_at-k_a\frac{L}{2}+\phi)}+\frac{\Delta \vec{X}_a}{2}\left(e^{-i(\omega_+t-k_+\frac{L}{2}+\phi_+)}+e^{-i(\omega_-t-k_-\frac{L}{2}+\phi_-)}\right)\right)\right]\, , 
\end{align}
\end{widetext}
the additional phase $e^{ik_aL}$ shows the time delay of $E^1$ compared to $E^0$ after half a round trip. This occurs several times and after an infinite number of round trips N, the full contribution of the external applied field inside the cavity is
\begin{widetext}
\begin{subequations}
\begin{align}
&\vec E^\mathrm{tot}_a(x=0,t) = \sum_{n=0}^{N=+\infty} \vec E^{n}_a(x=0,t) \,\\
&=t\vec X_a \Re \left[e^{-i(\omega_at+\phi)}\frac{e^{i\frac{k_aL}{2}}}{1+re^{ik_aL}}\right] + t\Delta \vec X_a\Re \left[e^{-i(\omega_+t+\phi_+)}\frac{e^{i\frac{k_+L}{2}}}{1+re^{ik_+L}}\right] + t\Delta \vec X_a\Re \left[e^{-i(\omega_-t+\phi_-)}\frac{e^{i\frac{k_-L}{2}}}{1+re^{ik_-L}}\right]\, \\
&\equiv \vec A(\omega_a)\cos(\omega_at+\phi)+\vec B(\omega_a)\sin(\omega_at+\phi) + \frac{\Delta X_a}{2 X_a}\sum_{i=\pm}\left(\vec A(\omega_i )\cos(\omega_it+\phi_i)+\vec B(\omega_i )\sin(\omega_it+\phi_i)\right) \, ,\label{full_field_applied_simplified}
\end{align}
\end{subequations}
\end{widetext}
where we used the fact that $r < 1$ such that $r^N \rightarrow 0$ and with 
\begin{subequations}\label{eq:A_B}
\begin{align}
    \vec A(\omega_a) &= \frac{t\vec X_a \left(1+r\right)\cos\left(\frac{\omega_aL}{2c}\right)}{1+2r\cos(\frac{\omega_aL}{c})+r^2}\, , \\
    \vec B(\omega_a) &= \frac{t\vec X_a \left(1-r\right)\sin\left(\frac{\omega_aL}{2c}\right)}{1+2r\cos(\frac{\omega_aL}{c})+r^2}\, .
\end{align}
\end{subequations}
One can notice a constructive interference at the center of the cavity for even modes of the cavity, as expected.

\subsection{DM field}\label{ap:DP_field}

Starting from the DM induced electric field with unknown polarization direction $\vec X_\mathrm{DM}$ and angular frequency $\omega$, the same procedure as above can be realized to know the DM electric field amplitude at the center of the cavity. The subtleties of this calculation are that : 1) the field is emitted by the mirrors towards the center of the cavity, therefore the transmission coefficient $t$ factor is not present ; 2) only the DM polarization transverse to the mirror is reemitted, noted $\vec X_\mathrm{DM,\parallel}$; and 3) there are two different contributions, in phase, each being emitted from one of the edges of the cavity. The total DM contribution at the center is then
\begin{widetext}
\begin{subequations}
\begin{align}
&\vec E^\mathrm{tot}_\mathrm{DM}(x=0,t) = \Re \left[\vec X_\mathrm{DM}e^{-i\omega t} + 2 \vec{X}_\mathrm{DM,\parallel}e^{-i(\omega t-\frac{kL}{2})}\frac{1}{1+re^{ikL}}\right]\, \label{full_field_DM} \\
&\equiv \vec C(\omega)\cos(\omega t)+\vec D(\omega)\sin(\omega t) \, , \label{full_field_DM_simplified}
\end{align}
\end{subequations}
\end{widetext} 
with
\begin{subequations}\label{eq:C_D}
\begin{align}
    \vec C(\omega) &= \vec X_\mathrm{DM} +\frac{2\vec X_\mathrm{DM,\parallel} (1+r)\cos(\frac{\omega L}{2c})}{1+2r\cos(\frac{\omega L}{c})+r^2}\, , \\
    \vec D(\omega) &= \frac{2\vec X_\mathrm{DM,\parallel}(1-r)\sin(\frac{\omega L}{2c})}{1+2r\cos(\frac{\omega L}{c})+r^2}\, .
\end{align}
\end{subequations}
The first term of Eq.~\eqref{full_field_DM} corresponds to the background oscillating DM field at the center, which is always present, even without cavity. The second term is the DM contribution from the cavity, which is almost equivalent, in its form, to the total contribution of the applied field, with an additional factor two, due to the emission of a field from both edges of the cavity (instead of only one for the applied field). 

\section{\label{ap:signal_noise_amp_annex}Signal and noise amplitudes}

In this section, we wish to derive the expressions of the signal contribution inside the cavity in Eq.~\eqref{eq:Etot2} and of the noise in Eq.~\eqref{general_E_power_noise}.

\subsection{Signal contribution}\label{ap:signal_amp}

From Eq.~\eqref{general_E_power}, and using Eqs.~\eqref{eq:A_B},\eqref{eq:C_D}, we can write the signal contribution as
\begin{widetext}
\begin{subequations}
\begin{align}
  P(\omega,\omega_a) &=  \sqrt{\left(\vec A(\omega_a)\cdot \vec C(\omega)+\vec B(\omega_a)\cdot \vec D(\omega)\right)^2+{\left(\vec B(\omega_a)\cdot \vec C(\omega)-\vec A(\omega_a)\cdot \vec D(\omega)\right)^2}}\,\\
    &= X_a X_\mathrm{DM}\beta\sqrt{\left( A'(\omega_a)^2+B'(\omega_a)^2\right)\times\left(C'(\omega)^2+ D'(\omega)^2\right)}
\end{align}
\end{subequations}
\end{widetext}
where we used $\vec X_a \cdot \vec X_\mathrm{DM} = \vec X_a \cdot \vec X_\mathrm{DM,\parallel} = X_a X_\mathrm{DM} \beta$ and where the prime quantities are defined as  
\begin{subequations}
\begin{align}
A'(\omega_a) &\equiv \frac{t\left(1+r\right)\cos\left(\frac{\omega_aL}{2c}\right)}{1+2r\cos(\frac{\omega_aL}{c})+r^2}\, , \\
B'(\omega_a) &\equiv \frac{t\left(1-r\right)\sin\left(\frac{\omega_aL}{2c}\right)}{1+2r\cos(\frac{\omega_aL}{c})+r^2}\, , \\
C'(\omega) &\equiv 1 +\frac{2(1+r)\cos(\frac{\omega L}{2c})}{1+2r\cos(\frac{\omega L}{c})+r^2}\, , \\
D'(\omega) &\equiv \frac{2(1-r)\sin(\frac{\omega L}{2c})}{1+2r\cos(\frac{\omega L}{c})+r^2}\, ,
\end{align}
\end{subequations}
i.e the polarizations are factorized from the amplitude functions $\{\vec A, \vec B, \vec C, \vec D\}$. The signal amplitude can be easily simplified to 
\begin{widetext}
\begin{subequations}
    \begin{align}
       P(\omega,\omega_a) = \frac{t X_aX_\mathrm{DM}\beta}{\sqrt{1+2r\cos(\frac{\omega_a L}{c})+r^2}}\sqrt{1+4\frac{1+(1+r)\cos(\frac{\omega L}{2c})}{1+2r\cos(\frac{\omega L}{c})+r^2}}\, ,
    \end{align}
    \end{subequations}
\end{widetext}
and $X_\mathrm{DM} = \chi c \sqrt{2\mu_0 \rho_\mathrm{DM}}$ from Eq.~\eqref{amp_E_field}.

\subsection{Noise contribution}\label{ap:syst_noise_amp}

Starting from the full expression of the systematic effect Eq.~\eqref{full_E_power_noise}, we can derive the noise amplitude amplified by the cavity as the sum of three different contributions, i.e 
\begin{widetext}
\begin{subequations}
\begin{align}
\mathrm{N}_1(\omega,\omega_a) &\equiv A(\omega_a)^2\left(\left(A(2\omega_a-\omega)+A(\omega)\right)^2+\left(B(2\omega_a-\omega)-B(\omega)\right)^2\right)\, , \\
\mathrm{N}_2(\omega,\omega_a) &\equiv B(\omega_a)^2\left(\left(A(2\omega_a-\omega)-A(\omega)\right)^2+\left(B(2\omega_a-\omega)+B(\omega)\right)^2\right)\, , \\
\mathrm{N}_3(\omega,\omega_a) &\equiv 4A(\omega_a)B(\omega_a)\left(A(2\omega_a-\omega)B(\omega)+B(2\omega_a-\omega)A(\omega)\right)\, , \\
\mathrm{N}(\omega,\omega_a) &\equiv \mathrm{N}_1(\omega,\omega_a)+\mathrm{N}_2(\omega,\omega_a)+\mathrm{N}_3(\omega,\omega_a)\, .
\label{noise_amp}
\end{align}
\end{subequations}
\end{widetext}
One can notice $N_i(\omega)$'s, and by extension $N(\omega)$, are of order 4 in $X_a$, making $\sqrt{N(\omega,\omega_a)} \propto X^2_a$, as expected.

\bibliographystyle{apsrev4-1}
\bibliography{DP_paper} 

\end{document}